\documentclass[11pt]{article}

\usepackage{fullpage}
\usepackage{graphicx}
\usepackage{xcolor}
\usepackage{url}

\begin{document}

%
% The "title" command has an optional parameter, allowing the author to define a "short title" to be used in page headers.
\title{Thanks for Stopping By: A Study of “Thanks” Usage on Wikimedia}

%
% The "author" command and its associated commands are used to define the authors and their affiliations.
% Of note is the shared affiliation of the first two authors, and the "authornote" and "authornotemark" commands
% used to denote shared contribution to the research.
\author{Swati Goel
\\Henry M. Gunn High School
\\ms.swati.goel@gmail.com
\and
Ashton Anderson
\\University of Toronto
\\ashton@cs.toronto.edu
\and
Leila Zia
\\Wikimedia Foundation
\\leila@wikimedia.org}

\newcommand{\swati}[1]{{\textcolor{green}{[SG: #1]}}}
\newcommand{\ashton}[1]{{\textcolor{blue}{[AA: #1]}}}
\newcommand{\leila}[1]{{\textcolor{brown}{[LZ: #1]}}}
\newcommand{\ttf}[1]{#1}
\newcommand{\thnx}[1]{``#1''}
\newcommand{\thnxtab}[1]{#1}
\newcommand{\tgr}[1]{#1}
\newcommand{\engt}[1]{#1}
\newcommand{\engthnked}[1]{#1}

%
% By default, the full list of authors will be used in the page headers. Often, this list is too long, and will overlap
% other information printed in the page headers. This command allows the author to define a more concise list
% of authors' names for this purpose.

%
% The abstract is a short summary of the work to be presented in the article.

\maketitle

\begin{abstract}
The \ttf{Thanks feature} on Wikipedia, also known as \thnx{Thanks}, is a tool with which editors can quickly and easily send one other positive feedback~\cite{florin2018ThanksWithURL}. The aim of this project is to better understand this feature: its scope, the characteristics of a typical \thnx{Thanks} interaction, and the effects of receiving a \tgr{thank} on individual editors. We study the motivational impacts of \thnx{Thanks} because maintaining editor engagement is a central problem for crowdsourced repositories of knowledge such as Wikimedia. Our main findings are that most editors have not been exposed to the \ttf{Thanks feature} (meaning they have never given nor received a \tgr{thank}), \tgr{thanks} are typically sent upwards (from less experienced to more experienced editors), and receiving a \tgr{thank} is correlated with having high levels of editor engagement. Though the prevalence of \thnx{Thanks} usage varies by editor experience, the impact of receiving a \tgr{thank} seems mostly consistent for all users. We empirically demonstrate that receiving a \tgr{thank} has a strong positive effect on short-term editor activity across the board and provide preliminary evidence that \tgr{thanks} could compound to have long-term effects as well.
\end{abstract}

\section{Introduction}
Wikipedia is an online encyclopedia that functions largely because of volunteer editors. Editor engagement and motivation are therefore central to Wikipedia’s existence. Research has long shown that positive external motivation (rewards, recognition) can lead to an increase in contribution to a community, and it is well-established that people will contribute more to a group the more enjoyable they find it~\cite{DBLP:journals/jcmc/LingBLWCLCFTRRK05}. A positive community environment, which provides editors a better experience, can therefore increase editor activity. A positive environment may actually be one of the most crucial elements for increasing engagement, as social factors tend to outweigh even those surrounding usability with regards to positively affecting contribution~\cite{DBLP:conf/chi/LampeWVO10}. The impact of these social factors could be quite significant, as a community member's internal value systems can be influenced by external rewards, thus making positive feedback an extremely useful tool in building online communities~\cite{DBLP:conf/hicss/TedjamuliaDOA05}. The \ttf{Thanks feature} could therefore represent an important resource for building a positive Wiki community.

\thnx{Thanks} is no longer a new Wiki feature, having been implemented on English Wikipedia on May 30th, 2013 and introduced to all projects soon thereafter. In contrast to previously existing features such as WikiLove, it allows editors to \engt{thank} one another for {\em specific} revisions in a {\em semi-private} way. There is a public record of who \engthnked{thanked} whom and when, but the specific revision and the article for which a user was \engthnked{thanked} is private information. This setup is quite different from that of previous community building features such as WikiLove, yet the literature on \thnx{Thanks} is limited. Previous research on \thnx{Thanks} includes a study by Harburg and Matias~\cite{harburgmatias2014AppreciationWithURL}, which draws an interesting contrast between the \ttf{Thanks feature} and WikiLove. We do not interact directly with their work as their main focus is analyzing usage differences between the two features whereas we study the impact of \thnx{Thanks} as well as its usage more deeply. Part of our work builds upon the thoughtful cross-cultural analysis of Nemoto and Okada~\cite{DBLP:journals/corr/NemotoO15}, which examines differences in \thnx{Thanks} usage across languages. In addition, we conduct more in-depth research on how \thnx{Thanks} is used over different levels of editor experience and how \tgr{thanks} are distributed over time. We also examine the \ttf{Thanks feature’s} impact on editor engagement, controlling for editor experience between comparisons of \engthnked{thanked} and \engthnked{unthanked} editors, and we find that \tgr{thanks} are linked to increases in short-term activity, in contrast to the results of the Nemoto-Okada study. This is significant because, in conjunction with our findings of the \ttf{Thanks feature’s} limited scope, our results suggest that \thnx{Thanks} has additional potential to increase editor motivation.
\section{Methodology and Results}

\begin{table}[tp] % here, top, botton, separate page
\centering
\begin{tabular}{|c||c|c|c|c|c|}
 \hline	 
Language & \thnxtab{Thanks} Givers & \thnxtab{Thanks} Receivers & Editors & \% \thnxtab{Thanks} Givers & \% \thnxtab{Thanks} Receivers\\
\hline
German & 23433 & 31567 & 390603 & 6.0 & 8.08 \\
\hline
Español & 14079 & 13924 & 526009 & 2.68 & 2.65\\
\hline
Italian & 8742 & 9412 & 186733 & 4.68 & 5.04\\
\hline
Portuguese & 8093 & 8593 & 194509 & 4.16 & 4.42\\
\hline
Polish & 5880 & 6506 & 83949 & 7.0 & 7.75\\
\hline
Farsi & 4611 & 4829 & 108114 & 4.26 & 4.47\\
\hline
Dutch & 4704 & 4830 & 89006 & 5.29 & 5.43\\
\hline
Arabic & 6873 & 6662 & 148112 & 4.64 & 4.5\\
\hline
Korean & 1855 & 1874 & 68285 & 2.72 & 2.74\\
\hline
Thai & 1045 & 610 & 29780 & 3.51 & 2.05\\
\hline
Norwegian & 1171 & 2500 & 41501 & 2.82 & 6.02\\
\hline

\end{tabular}
\caption{Scope of \ttf{Thanks feature} in select languages. Each column represents the absolute count of editors in a category unless explicitly specified as a percentage. English is excluded due to data restrictions.}
\label{tab:thanks_scope}
\end{table}
Our work is separated into two parts: analyzing how people interact with \thnx{Thanks} and assessing the impact the feature has on editor engagement and motivation. In part one, we present data on general use, such as the characteristics of the feature’s users, how likely different types of editors are to receive \tgr{thanks}, etc. We then establish a correlation between receiving a \tgr{thank} and having a higher edit count. This naturally leads to part two, in which we attempt to determine whether this link is causal. We discuss our results at a high-level below.

\subsection{How \thnx{Thanks} is Used}

\subsubsection{Scope of \thnx{Thanks}}

The scope of the \ttf{Thanks feature}, or the number of editors who have either given or received a \tgr{thank} since the feature was first introduced, is generally between 4-6\% (in a subset of larger languages), as shown in Table ~\ref{tab:thanks_scope}. In the set of editors with 5+ edits, the scope of the feature is 15-17\% (again in a subset of larger languages), indicating the existence of a small group of active editors who are responsible for a vast majority of \tgr{thanks}.

The \ttf{Thanks feature} is not as widely used as it could be, but it has become more prevalent in recent years, a trend shown in Table ~\ref{tab:thanks_over_time}. Even in languages where the absolute editor count has dropped, \thnx{Thanks} usage rates have increased. There is a clear upward trend in the number of \tgr{thanks} givers (relative to the number of total editors). Somewhat surprisingly, this same trend does not hold for \tgr{thanks} receivers. The generally increasing usage rates for givers suggest that some editors are only now beginning to use the \ttf{Thanks feature}, which could indicate that a large portion of the editor population remains unaware of it. If this is the case, efforts to increase exposure could have significant benefits, especially because novice editors—those for whom the feature could potentially be the most valuable—are currently the ones who interact with it the least.

\begin{table}[tp] % here, top, botton, separate page
\hspace{-0.3in}
\begin{tabular}{|c||c|c|c|c|c|c|c|c|}
 	\hline
	Language & Givers & Givers & \% Givers & \% Givers & Receivers & Receivers & \% Receivers & \% Receivers\\
  & 2018 & 2016 & 2018 & 2016 & 2018 & 2016 & 2018 & 2016\\
	\hline
 	Italian & 1910 & 1511 & 6.03 & 4.98 & 2195 & 1995 & 6.93 & 6.57\\
	\hline
	Portuguese & 1273 & 1314 & 5.16 & 4.75 & 1835 & 1601 & 7.44 & 5.79\\
\hline
Polish & 1297 & 940 & 8.67 & 6.66 & 1349 & 1419 & 9.02 & 10.06\\
\hline
Farsi & 1103 & 575 & 5.72 & 4.47 & 927 & 789 & 4.81 & 6.13\\
\hline
Netherlandic & 935 & 915 & 6.82 & 6.28 & 1046 & 1096 & 7.63 & 7.52\\
\hline
\end{tabular}
\caption{\thnx{Thanks} usage rates. Each column represents the absolute count of editors in a category unless explicitly specified as a percentage. Data is from the first 6 months of each year.}
\label{tab:thanks_over_time}
\end{table}

\subsubsection{Usage by Editor Experience}

The distribution of \tgr{thanks} shows that novice editors interact with the \ttf{Thanks feature} less frequently than their more experienced counterparts. Presented in Table ~\ref{tab:thanks_distribution} is the average number of \tgr{thanks} received by a set of editors (grouped by editor experience) per month or day counting only months or days in which they received at least one \tgr{thank}. We do not include editors who have never received a \tgr{thank}, and we partition the remaining data into novices (bottom 20\% of editors by edit count) and experienced (top 20\% of editors by edit count). The editors we studied received multiple \tgr{thanks} on the same day more often than they would have if \tgr{thanks} were given at random times, implying a non-uniform distribution, and there was a strong correlation between those who were given more \tgr{thanks} and those who had higher edit counts. These two findings taken together indicate that \tgr{thanks} are often received in ‘clumps' and awarded to those who are editing more frequently. This more concentrated \thnx{Thanks} usage may be why the feature has the outsized impact shown in part two, but it's also possible that a more widely used feature would retain the same benefits while reaching more editors.

\begin{table}[tp] % here, top, botton, separate page
\centering
\begin{tabular}{|c||c|c|c|c|}
 	\hline
	Language & Sample & Year & Month & Day\\
\hline
Italian & Bottom 20\% & 2.69 & 1.68 & 1.18\\
\hline
Italian & Top 20\% & 119.62 & 13.07 & 1.59\\
\hline
Portuguese & Bottom 20\% & 2.95 & 1.98 & 1.34\\
\hline
Portuguese & Top 20\% & 206.24 & 22.22 & 2.33\\
\hline
Polish & Bottom 20\% & 2.34 & 1.63 & 1.19\\
\hline
Polish & Top 20\% & 48.63 & 6.3 & 1.42\\
\hline
Farsi & Bottom 20\% & 2.73 & 1.91 & 1.28\\
\hline
Farsi & Top 20\% & 123.0 & 13.74 & 1.74\\
\hline
Dutch & Bottom 20\% & 2.37 & 1.48 & 1.11\\
\hline
Dutch & Top 20\% & 81.0 & 9.61 & 1.5\\
\hline
\end{tabular}
\caption{Distribution of \tgr{thanks} for individual editors. Each number represents the average number of thanks received by an editor in a timeframe considering only those in which at least one thank was received.}
\label{tab:thanks_distribution}
\end{table}

\begin{figure}[htbp]
  \centering
  \includegraphics[width=4in]{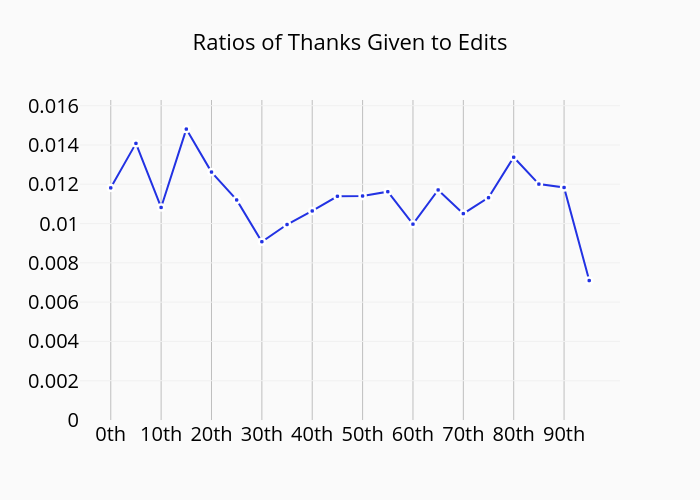}
  \caption{\thnx{Thanks} by edit count ratios. Percentile are based on edit count. Data from June '17-'18}
  \label{fig:edit_count_ratios}
\end{figure}

The existence of a positive \tgr{thanks} to edit count correlation is further corroborated by our analysis of average \tgr{thanks} given over all editor percentiles: the top 5\% of editors give by far the most \tgr{thanks} in absolute terms. However, as Figure~\ref{fig:edit_count_ratios} shows, these editors give the least \tgr{thanks} relative to their edit counts.

\subsubsection{Additional Work}

The research page for this project~\cite{ziagoel2018ThanksWithURL} contains links to a number of sub-projects left out of this report as well as code pipelines for anyone interested in replicating the project. A brief summary of some of these projects follows:
\begin{itemize}
\item In order to better understand the scope of \thnx{Thanks}, we define different levels of editor engagement with the feature. We calculate the percentage of total editors who contributed to random samplings of \tgr{thanks} and find that just under 5\% of editors in our dataset (a group of larger Wikipedia languages) were responsible for giving 80\% of \tgr{thanks}.
\item To characterize \tgr{thanks} senders vs \tgr{thanks} receivers, we compare the average tenure and edit count of the former group to the average tenure and edit count of the latter group and find that in most languages studied—Norwegian being a notable exception—\tgr{thanks} are generally sent from newer, less experienced editors to more experienced, veteran editors. This trend is further corroborated by Nemoto and Okada.
\item To get a sense of the variance in \thnx{Thanks} usage across projects, we rank all Wikimedia projects at the time of study (July 2018) by the ratio of editors who had sent a \tgr{thank} to the total number of editors since \thnx{Thanks} was introduced.
\end{itemize}

\begin{table}[tp] % here, top, botton, separate page
\centering
\begin{tabular}{|c||c|c|c|c|c|c|c|}
 	\hline
	{\bf Group} & Tenure & Edits & Thanks & Short-term & Short-term & Next Day's
& Editors with\\
  & & & & Edits & Thanks & Edits
& Higher Counts\\
\hline
\engthnked{Thanked} & 2848.3 & 555.3 & 5.8 & 56.8 & 0.3 & 9.6 & 46\\
\hline
\engthnked{Unthanked} & 3021.7 & 573.1 & 5.9 & 60.6 & 0.3 & 5.7 & 15\\
\hline
\end{tabular}
\caption{Motivation study results sample}
\label{tab:motivation}
\end{table}

\subsection{Editor Engagement}
In part two, we establish a link between receiving a \tgr{thank} and having a higher future edit count, at least in the short-term. To do this, we match editors who received a \tgr{thank} on some day with editors who did not receive a \tgr{thank} on some (potentially different) day and compare their subsequent edit activity. Because we only match between editors with similar characteristics, we can be reasonably confident that the \tgr{thank}, and not some other factor, causes the future edit count differences we see.
Table~\ref{tab:motivation} presents a result of the study in Polish Wikipedia (trials in Portuguese Wikipedia and MetaWiki yielded similar results). In the table, the feature data for each cohort (ex: tenure for \engthnked{thanked} editors) is an average of that feature's value over all members of that group. The ‘editors with higher counts’ field represents the number of editors of the group who had a higher subsequent edit count than their match.

We find a positive correlation between the five features we use (tenure, edits, \tgr{thanks}, short-term edits, short-term \tgr{thanks}) and the dependent variable, future edit count. In Table ~\ref{tab:motivation}, \engthnked{unthanked} editors have a higher average value for every feature. We would therefore expect them to have a higher average future edit count as well. This is not reflected in the data, suggesting that the difference we see in future edit count is caused by the one field along which the groups are not balanced: whether or not an editor has just been \engthnked{thanked}.
It is possible that the edit count discrepancy is actually caused by some unaccounted for confounding variable, but our test results suggest this is not the case. A random forest classifier we trained, for example, demonstrated that the features we chose have good predictive accuracy (a little over 90\%). Using both the features and a field for whether the treatment of receiving a \tgr{thank} was applied, the classifier accurately determined the range in which future edit count would fall. This suggests that our features were comprehensive. Additionally, another test revealed that a variety of random features either had less weight on the overall prediction than our selected five or led to similar matchings.

\section{Conclusions}
\thnx{Thanks} usage rates have been increasing over time, even in projects where the absolute editor count has dropped. This could mean that current low \thnx{Thanks} usage rates among novice editors are due to editors being unaware of the feature as opposed to them being opposed to it, and it indicates that increasing the \ttf{Thanks feature’s} exposure could increase usage. Currently, \tgr{thanks} seem to be centered around a few editors and are not sent as a matter of course, which could change with increased attention to the feature. More attention might also alter the \thnx{Thanks} social structure in which most \tgr{thanks} are sent from less experienced to more experienced editors. Having more positive feedback go from experienced editors to novices seems intuitively likely to encourage newer editors to stay involved, and in fact, we show that \thnx{Thanks} is linked to higher short-term editor activity. Because the feature is largely used by a group of editors who are already active and committed, each individual \tgr{thank} is unlikely to have more than a short-term impact. However, given our results, it's conceivable that the effects of \tgr{thanks} may compound over time and that receiving a \tgr{thank} as a novice editor could change a Wikimedian’s career. While the long-term effects of \tgr{thanks} were not determined in this study, our findings suggest that increasing \thnx{Thanks} usage would positively affect editor retention and activity. We would not want \tgr{thanks} to become so common as to be meaningless, but the feature is far from that point, if it exists. Thus, it is our belief that connecting more editors with the \ttf{Thanks feature} would be beneficial to editor motivation and possibly editor retention.

\bibliographystyle{plain}
\bibliography{thanks}

\end{document}